\documentclass[aps,twocolumn]{revtex4}
\usepackage{amsfonts}
\usepackage{amsmath}
\usepackage{mathrsfs}
\usepackage{amssymb}
\usepackage{graphicx}
\usepackage{dcolumn}
\usepackage{bm}
\usepackage{color}
\usepackage[colorlinks,linkcolor=blue,anchorcolor=blue,citecolor=blue,urlcolor=black]{hyperref}

\begin{document}

\title{Optomechanically Induced Transparency at Exceptional Points}

\author{Hao L\"{u}$^{1,2,4}$, Changqing Wang$^3$, Lan Yang$^3$, and Hui Jing$^{2,}$}
\email[]{jinghui73@foxmail.com}

\affiliation{$^1$Key Laboratory for Quantum Optics, Shanghai Institute of Optics and Fine Mechanics, Chinese Academy of Sciences, Shanghai 201800, China}
\affiliation{$^2$Department of Electrical and Systems Engineering, Washington University, St. Louis, Missouri 63130, USA}
\affiliation{$^3$Key Laboratory of Low-Dimensional Quantum Structures and Quantum Control of Ministry of Education, Department of Physics and Synergetic Innovation Center for Quantum Effects and Applications, Hunan Normal University, Changsha 410081, China}
\affiliation{$^4$University of Chinese Academy of Sciences, Beijing 100049, China}

\begin{abstract}

We study optomechanically induced transparency in a microresonator coupled with nanoparticles. By tuning the relative angle of nanoparticles, exceptional points (EPs) emerge periodically in this system and thus strongly modify both the transmission rate and the group delay of the signal. As a result, controllable slow-to-fast light switch can be achieved by manipulating external nanoparticles. This provides a new way to engineer EP-assisted optomechanical devices for applications in optical communications and signal processing.

\end{abstract}

\date{\today}

\maketitle

\section{Introduction}

Cavity optomechanics (COM), with rapid advances in the last decade \cite{AKM2014,Metcalfe2014,Midolo2018,Bowen2017}, has led to many important applications, such as optoelectronic quantum transducer \cite{Polzik2014,Teufel2016}, ground-state cooling of motion \cite{Chan2011,Teufel2011}, phonon lasing or squeezing \cite{Wollman2015,Lei2016,Vahala2010,Jing2014}, and weak-force sensing \cite{Arcizet2006,Kippenberg2012}. Optomechanically induced transparency (OMIT), as an example closely relevant to COM, provides a new strategy for the realization of coherent control or even the quantum memory of light, which can be broadly applied to numerous physical platforms, including solid-state devices \cite{Weis2010,Safavi-Naeini2011,Kippenberg2013,Vitali2013,Shen,Xiao2016}, atomic gases \cite{EIT2005}, and even liquid droplets \cite{Carmon2018}. In view of the wide range of COM devices achieved nowadays, OMIT provides a versatile platform to explore exotic effects such as nonlinear or cascaded OMIT effects \cite{Jing2015,Nonlinear,Liu,Cascade,Xiong2012,Jing2016} and nonreciprocal physics \cite{Dong2016,Painter2017,spinning}, leading to many applications such as wavelength conversion \cite{Painter2012} and highly sensitive sensors \cite{Wu2017}.

In parallel, properties and applications of exceptional-point (EP) systems, especially EP optics, have attracted intense interests in recent years \cite{Bender1998,Bender2007,PT2016,Feng2017PT,Peng2014,Ruter2014,Zyablovsky,Longhi2018}. In such systems, two or more eigenmodes coalesce at the EPs, leading to a variety of unconventional effects observed in experiments, such as loss-induced coherence \cite{Guo2009,Peng2014Science}, unidirectional lasing \cite{Feng2017}, invisible sensing \cite{Invisibility}, robust wireless power transfer \cite{Fan2017}, and exotic topological states \cite{Zhen2018,YangBiao2018}.
EP effects in COM have also been probed both theoretically and experimentally \cite{Jing2014,Harris2016,Jing2017,Jing2017cooling}, such as low-power phonon laser \cite{Jing2014,Jing2017}, high-order EPs in COM \cite{Jing2017cooling}, and nonreciprocal COM devices \cite{Harris2016,Renault2017}, highlighting new opportunities of  enhancing or steering coherent light-matter interactions with the new tool of EPs.

Very recently, by coupling a whispering-gallery-mode (WGM) microresonator with two external nanoparticles, periodic emergence of EPs has been observed experimentally when tuning the relative positions of the particles \cite{Yang2016}. Counterintuitive EP effects, such as modal chirality \cite{Yang2016} and highly-sensitive sensing \cite{Yang2017}, have been revealed in these exquisite devices. On the basis of these experiments, here we study the new possibility of COM control by tuning the relative positions of nanoparticles along the circumference of the resonator. We find that due to the particle-induced asymmetric backscattering between two near-degenerate resonator modes \cite{Wiersig2014,Wiersig2016}, the OMIT spectrum is strongly modified in the vicinity of EPs. In particular, a slow-to-fast light switch can be achieved by tuning the relative angle of the particles, which is of practical use in, for example, optical signal processing and communications. Our work, without the need for any optical gain or complicated refractive-index modulation \cite{Yang2016,Yang2017}, is well within current experimental abilities. More studies on, for example, nonlinear or topological COM \cite{Clerk2016,Verhagen2017,Marquardt2015} can also be envisaged with this single-resonator EP device.

\begin{figure*}[ht]
\centering
\includegraphics[width=6.2in]{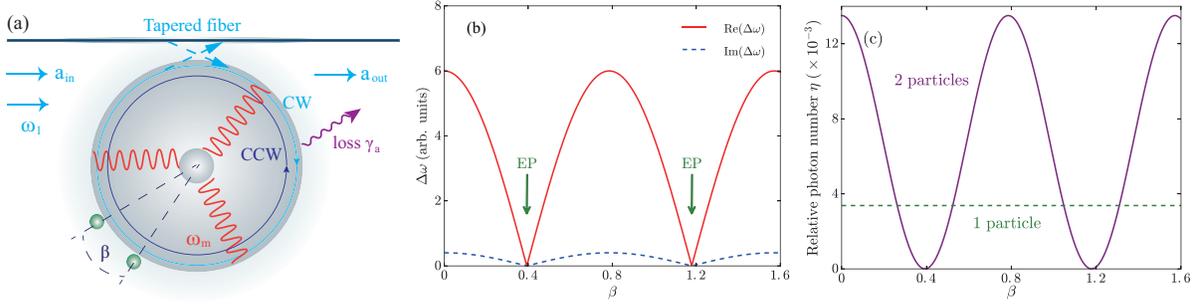}
\caption{(a) Optomechanics in a microresonator with two nanoparticles in the WGM field. The resonator supports a mechanical mode at frequency $\omega_m$, which is driven by a pump field at frequency $\omega_l$. $\beta$ is the relative angle between the two particles. The complex frequency splitting $\Delta\omega$ (b) and the relative photon number $\eta$ £š(c) as a function of the angle $\beta$. In (c), we choose $P_l=1\,$mW.}
\label{fig1}
\end{figure*}

\section{Theoretical model}

As in recent experiments \cite{Yang2016,Yang2017}, we consider a WGM microresonator (of resonance frequency $\omega_a$ and damping rate $\gamma_a$) coupled with two nanoparticles (see Fig.\,\ref{fig1}). This resonator is driven by a strong pump field at frequency $\omega_l$ and a weak probe field at frequency $\omega_p$. The amplitudes of the pump and probe fields are
$$E_l=\sqrt{P_l /\hbar\omega_l},~~~E_p=\sqrt{P_p/\hbar\omega_p},$$
where $P_l$ and $P_p$ denote the pump and probe power, respectively. The resonator also supports a phonon mode with the mechanical frequency $\omega_m$ and damping rate $\gamma_m$. Two silica nanotips as Rayleigh scatterers, which can be fabricated by the wet etching of tapered fiber tips prepared by heating and stretching standard optical fibers \cite{Yang2016,Yang2010}, are placed in the evanescent field of the resonator. The position of each particle can be controlled by a nanopositioner, which tunes the relative position and effective size of the nanotip in the WGM fields. The non-Hermitian optical coupling of the clockwise- (CW) and counterclockwise- (CCW) traveling waves, induced by the nanoparticles [54], can be described
by the scattering rate \cite{Yang2017}
$$J_{1,2}=\epsilon_1 + \epsilon_2 e^{\mp i2m\beta},$$
which corresponds, respectively, to the scattering from the CCW (CW) to CW (CCW) modes. Here, $m$ is the azimuthal mode number, $\beta$ is the relative angular position of the two scatterers, and $\epsilon_j$ ($j=1,2$) is half of the complex frequency splitting induced by the $j$th scatterer. Steering of the angle $\beta$ can bring the system to EPs, as already observed experimentally \cite{Yang2016,Yang2017}.
The purpose of our work here is to show the role of EPs, achieved by tuning $\beta$ \cite{Yang2016,Yang2017}, in the OMIT process.

In the rotating frame at the pump frequency, the total Hamiltonian of the system can be written at the simplest level as
\begin{align}
H=&H_0+H_m+H_{\mathrm{int}}+H_{\mathrm{dr}},\nonumber\\
H_0=&\Delta a^{\dagger}_{\rm cw}a_{\rm cw}  + \Delta a^{\dagger}_{\rm ccw}a_{\rm ccw}+ J_1 a^{\dagger}_{\rm cw}a_{\rm ccw} + J_2 a^{\dagger}_{\rm ccw}a_{\rm cw}, \nonumber\\
H_m=&\frac{p^2}{2m_{\rm eff}} + \frac{1}{2}m_{\rm eff}\omega^2_m x^2,\nonumber\\
H_{\mathrm{int}}=&-g x(a^{\dagger}_{\rm cw}a_{\rm cw}+ a^{\dagger}_{\rm ccw}a_{\rm ccw}),\nonumber\\
H_{\mathrm{dr}}=&i\sqrt{\gamma_{\rm ex}}\left[(E_l+E_p e^{-i\xi t})a^{\dagger}_{\rm cw} -h.c.\right],
\label{eq:H}
\end{align}
where $\xi=\omega_p-\omega_l$ and
$$\Delta=\Delta_a+\mathrm{Re}(\epsilon_1+\epsilon_2)~~~\Delta_a=\omega_a-\omega_l.$$
In consideration of a small change of the cavity length, the COM coupling coefficient can be written as $g=\omega_a/R$ \cite{Vahala2010}, where $R$ is the radius of the resonator. $\gamma_{\mathrm{ex}}$ is the loss induced by the resonator-fiber coupling, $m_{\rm eff}$ is the effective mass of the mechanical mode, $a_{\mathrm{cw,ccw}}$ are the optical annihilation operators for the CW and CCW modes, and $x$ and $p$ denote the displacement and momentum operators, respectively.

The nanoparticles induce frequency splitting of the optical modes, and the corresponding eigenfrequencies can be derived as
\begin{eqnarray}
\omega_{1,2}&=&\omega_a-i\gamma_a+\epsilon_1+\epsilon_2 \nonumber\\
&&\pm\sqrt{\epsilon^2_1+\epsilon^2_2+2\epsilon_1\epsilon_2\cos(2m\beta)}.
\end{eqnarray}
Since $\epsilon_{1,2}$ are complex numbers, the complex eigenvalue splitting $\Delta\omega$ is allowed to be zero by tuning the angle $\beta$, which brings the system to an EP. For $J_1=J_2=0$, i.e., when the two scatterers are absent, $H_0$ has two orthogonal eigenstates with the same frequency. For $J_1\neq J_2$ and $J_1J_2=0$, $H_0$ has only one eigenvalue and one eigenvector, indicating the emergence of an EP \cite{Wiersig2014,Wiersig2016}. In this case, the critical value of $\beta$ can be obtained as
\begin{align}
\beta_{c}=\frac{l\pi}{2m}\mp\frac{\arg{(\epsilon_1)}-\arg{(\epsilon_2)}}{2m}~~~(l=\pm1,\pm3,...),
\end{align}
where $\mp$ corresponds to $J_1=0$ or $J_2=0$. Here, $|\epsilon_1|=|\epsilon_2|$, required for the realization of EPs, can be achieved in experiments by tuning the distance between the resonator and the particles \cite{Yang2016,Yang2017}. By continuously tuning $\beta$, $\beta_c$ emerges periodically, as demonstrated in recent experiments \cite{Yang2016,Yang2017}. We note that the particles are not necessarily identical, and EPs can be achieved in the situation in which two particles have a tiny difference in their electric permittivity. In fact, as already demonstrated in previous works \cite{Yang2016}, this can be achieved by adjusting the relative position of the scatterers and the resonator, so that each scatterer has the same spatial overlap with the evanescent field \cite{Yang2016}. Figure\,\ref{fig1}(b) shows the frequency splitting $\Delta\omega$ as a function of $\beta$, in which we choose $\epsilon_1/\gamma_a = 1.5-0.1i$, $\epsilon_2/\gamma_a = 1.4999-0.1015i$, and $m=4$ \cite{Yang2016}. We find that, in good agreement with the experiments, the real and imaginary parts of $\Delta\omega$ oscillate with the same period \cite{Yang2016,Yang2017}. The EPs correspond to the points at which $\Delta\omega = 0$, (e.g., for $\beta_ c = 0.3926$, we have $J_1=0$ and $J_2/\gamma_a=0.0002+0.003i$). In the following, we focus on the exotic features of OMIT in the vicinities of EPs, including the transmission rate and the group delay of the signal.

\begin{figure*}[ht]
\centering
\includegraphics[width=5in]{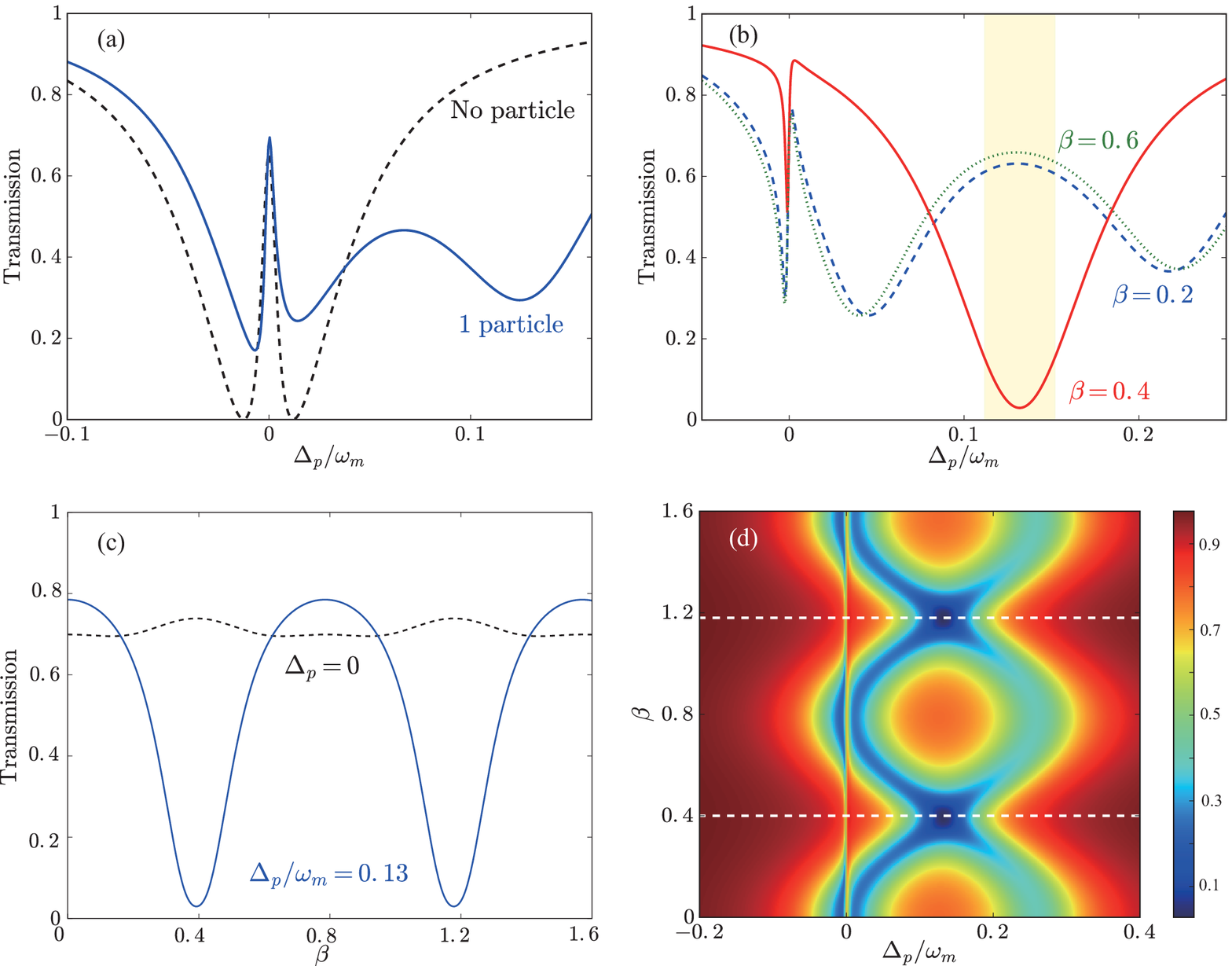}
\caption{Transmission of the probe light as a function of the optical detuning $\Delta_p$ (a,b) and the relative angle $\beta$ (c). (d) Transmission as a function of the optical detuning $\Delta_p$	and the angle $\beta$. In all of these figures, we have selected $P_l=1\,$mW and $\omega_a/\omega_m=1$.}
\label{fig2}
\end{figure*}

For this purpose, we write down the equations of motion of the system, i.e.,
\begin{eqnarray}
\dot{a}_{\rm cw}&=&-(i\Delta-ig x+\gamma)a_{\rm cw}-iJ_1a_{\rm ccw}+\sqrt{\gamma_{\rm ex}}E_l\nonumber\\
&&+\sqrt{\gamma_{\rm ex}}E_p e^{-i\xi t},\nonumber\\
\dot{a}_{\rm ccw}&=&-(i\Delta-ig x+\gamma)a_{\rm ccw}-iJ_2a_{\rm cw},\nonumber\\
\ddot{x}&=& -\gamma_m\dot{x}-\omega^2_m x+ \frac{g}{m_{\mathrm{eff}}}(a^{\dagger}_{\rm cw}a_{\rm cw}+a^{\dagger}_{\rm ccw}a_{\rm ccw}),
\label{eq:dy}
\end{eqnarray}
where $\gamma=\gamma_a-\mathrm{Im}(\epsilon_1+\epsilon_2)$ is the total optical loss, from which we obtain the steady-state values of the dynamical variables as
\begin{align}
\bar{a}_{\rm cw}=&\frac{\sqrt{\gamma_{\rm ex}}E_l(\gamma+i\Delta-ig\bar{x})}{(\gamma+i\Delta-ig\bar{x})^2+J_1J_2},\nonumber\\
\bar{a}_{\rm ccw}=&\frac{-i\sqrt{\gamma_{\rm ex}}E_l J_2}{(\gamma+i\Delta-ig\bar{x})^2+J_1J_2},\nonumber\\
\bar{x}=&\frac{g\gamma_{ex}|E_l|^2\left[|J_2|^2+\gamma^2+(\Delta-g\bar{x})^2 \right] }{m_{\mathrm{eff}}\omega^2_m |(\gamma+i\Delta-ig\bar{x})^2+J_1J_2|^2}.
\end{align}
We see that the $\beta$-dependent optical coupling rate indeed affects both the intracavity optical intensity and the mechanical displacement. To see this more clearly, we define the relative photon number, as follows:
\begin{align}
\eta=\frac{|\bar{a}_{\rm ccw}|^2}{|\bar{a}_{\rm cw}|^2}=\left|\frac{J_2}{\gamma+i\Delta-i\xi\bar{x}} \right|^2.
\end{align}
For the single-particle case, $\eta$ is a constant; however, when another nanoparticle is introduced, this situation changes. In this case, $\eta$ is a periodic function of $\beta$ [see Fig.\,\ref{fig1}(c)]. The CCW mode is absent around $\beta\approx 0.4$. At this point, we have $J_2=0$, indicating that the scattering from the CW to CCW modes is prohibited. This means that the effective COM coupling or the mechanical oscillations can be tuned by the nanoparticles, which as a result, leads to strong modifications of OMIT.

To calculate the optical transmission rate, we expand every operator as the sum of its steady value and a
small fluctuation, i.e., $x=\bar{x}+\delta xe^{-i\xi t}+\delta x^{\ast}e^{i\xi t},$ and
\begin{equation}
a_{i(i=\rm cw,\rm ccw)}=\bar{a}_{i}+\delta a^{-}_{i}e^{-i\xi t}+\delta a^{+}_{i}e^{i\xi t}.
\label{eq:ansatz}
\end{equation}
Substituting Eqs. (\ref{eq:ansatz}) into Eqs. (\ref{eq:dy}) leads to
\begin{equation}
\chi^{-1}(\xi)\delta x - g\sum_{i=\rm cw}^{\rm ccw}(\bar{a}^{\ast}_{i}\delta a^{-}_{i}+
\bar{a}_{i}\delta a^{+\ast}_{i})=0,
\label{eq:dx}
\end{equation}
and
\begin{align}
f_1(\xi)\delta a^{-}_{\rm cw}+iJ_1\delta a^{-}_{\rm ccw}-ig\bar{a}_{\rm cw}\delta x=&\sqrt{\gamma_{\rm ex}}E_p,\nonumber\\
f_2(\xi)\delta a^{+\ast}_{\rm cw}-iJ^{\ast}_1\delta a^{+\ast}_{\rm ccw}+ig\bar{a}^{\ast}_{\rm cw}\delta x=&0,\nonumber\\
f_1(\xi)\delta a^{-}_{\rm ccw}+iJ_2\delta a^{-}_{\rm cw}-ig\bar{a}_{\rm ccw}\delta x=&0,\nonumber\\
f_2(\xi)\delta a^{+\ast}_{\rm ccw}-iJ^{\ast}_2\delta a^{+\ast}_{\rm cw}+ig\bar{a}^{\ast}_{\rm ccw}\delta x =&0,
\label{eq:side}
\end{align}
with $f_{1,2}(\xi)=\gamma\pm i(\Delta-ig\bar{x})-i\xi$ and
\begin{equation}
\chi^{-1}(\xi)=m_{\mathrm{eff}}(\omega_m^2-\xi^2-i\xi\gamma_m).
\end{equation}
By solving Eqs.\,(\ref{eq:dx}) and (\ref{eq:side}) we obtain
\begin{align}
\delta a^{-}_{\rm cw}=\frac{\sqrt{\gamma_{\rm ex}}E_p}{h_4(\xi)}\left[1 +\frac{\xi^2\chi(\xi)h_1(\xi)h_3(\xi)}{h_3(\xi)h_4(\xi)-\xi^2\chi(\xi)h(\xi) }
\right],
\end{align}
where,
\begin{align}
&h_1(\xi)=\left[i\bar{a}_{\rm cw}+J_1\bar{a}_{\rm ccw}/f_1(\xi)\right]\left[\bar{a}^{\ast}_{\rm cw}f_1(\xi)-i\bar{a}^{\ast}_{\rm ccw}J_2\right],\nonumber\\
&h_2(\xi)= \left[ i\bar{a}_{\rm ccw}f_1(\xi)+J_2\bar{a}_{\rm cw}\right] \left[ \bar{a}^{\ast}_{\rm cw}f_1(\xi)-i\bar{a}^{\ast}_{\rm ccw}J_2\right],\nonumber\\
&h_3(\xi)= f^2_2(\xi) + J^{\ast}_1J^{\ast}_2,\nonumber\\
&h_4(\xi)= f_1(\xi) + J_1J_2/f_1(\xi),\nonumber\\
&h_5(\xi)=J^{\ast}_1\bar{a}^{\ast}_{\rm ccw}\bar{a}_{\rm cw}+J^{\ast}_2\bar{a}^{\ast}_{\rm cw}\bar{a}_{\rm ccw}-i\bar{n}f_2(\xi) ,\nonumber\\
&h_6(\xi)= J_2\bar{a}^{\ast}_{\rm ccw}\bar{a}_{\rm cw}+J_1\bar{a}^{\ast}_{\rm cw}\bar{a}_{\rm ccw} + i\bar{n}f_1(\xi),\nonumber\\
&h(\xi)=h_3(\xi)h_6(\xi)+f_1(\xi)h_4(\xi)h_5(\xi),
\end{align}
with $\bar{n}=|\bar{a}_{\rm cw}|^2+|\bar{a}_{\rm ccw}|^2$. With these at hand, the expectation value of the output field can then be obtained by using the input-output relation
\begin{equation}
a_{\mathrm{out}}=a_{\mathrm{in}}- \sqrt{\gamma_{\mathrm{ex}}}\delta a^{-}_{\mathrm{cw}},
\end{equation}
where $a_{\mathrm{in}}$ and $a_{\mathrm{out}}$ are the input or output field operators, respectively. The transmission rate of the probe field is then
\begin{align}
T=\left|t_p\right|^2=\left|\frac{a_{\rm out}}{a_{\rm in}} \right|^2=\left|1-\frac{\sqrt{\gamma_{\mathrm{ex}}}}{E_p}\delta a^{-}_{\rm cw}\right|^2.
\end{align}
This sets the framework for our discussions of the impact of EPs on the transmission rate and the group delay of the
probe light.

\section{Results and discussions}

Figure\,\ref{fig2} shows the transmission rate as a function of the angle $\beta$ and the probe detuning $$\Delta_p=\omega_p-\omega_a.$$ 
Here, we have selected experimentally feasible values, i.e. $R=34.5\,\mu$m, $\omega_a=193\,$THz, $\gamma_a=\gamma_{\mathrm{ex}}=6.43\,$MHz, $m_{\mathrm{eff}}=50\,$ng, $\omega_m=147\,$MHz, $\gamma_m=0.24\,$MHz, $\epsilon_1/\gamma_a = 1.5 - 0.1i$, $\epsilon_2/\gamma_a = 1.4999 - 0.1015i$, and $m=4$ \cite{Yang2016,Peng2014}.
As in standard OMIT (without any particles), when the pump is detuned from the resonance by $\omega_m$, a single OMIT peak emerges around $\Delta_p=0$, due to the destructive interference of two absorption channels of the probe photons (by the resonator or by the phonon mode) \cite{Weis2010}. For completeness, we also plot the case with a single nanoparticle [see Fig.\,\ref{fig2}(a)], i.e., a deformed spectrum due to mode splittings of both the pump and the probe, which is similar to the case with coupled two resonators \cite{AKM2014,Jing2015}.

More interestingly, Fig.\,\ref{fig2}(b) shows the $\beta$-dependent transmission rate with two nanoparticles, featuring a Fano-like spectrum around the resonance, due to the interference between the probe and the scattered control field. For $\beta=0.2$ or 0.6, a transparency window emerges around $\Delta_p/\omega_m=0.13$, corresponding to the frequency shifts induced by the particles [see also Fig.\,\ref{fig1}(b)]. However, by tuning the system close to the EP (with $\beta=0.4$), strong absorption of the probe can be achieved; see the red curve in Fig.\,\ref{fig2}(b). Hence the relative angle between two nanoparticles can be steered to achieve not only the EPs but also the optical switching [see the highlighted zone in Fig.\,\ref{fig2}(b)]. In Fig.\,\ref{fig2}(c), we see that for $\Delta_p/\omega_m=0.13$, the transmission of the probe changes periodically with $\beta$, and particularly, the probe can be blocked at the EPs [see also Fig.\,\ref{fig2}(d)]. This could be of practical uses in making a passive EP device for optical engineering and communications.

\begin{figure}[ht]
\centering
\includegraphics[width=2.5in]{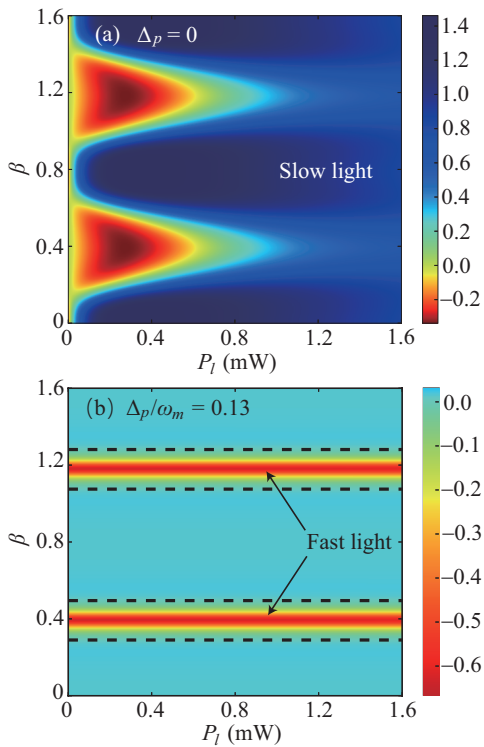}
\caption{(a) Group delay of the probe light $\tau_g$ (in units of $\mu$s) as a function of the pump power $P_l$ and angle $\beta$. (b) Group delay $\tau_g$ as a function of the pump power $P_l$. In both (a) and (b), we have selected $\Delta_a/\omega_m=1$.}
\label{fig3}
\end{figure}

\begin{figure}[ht]
\centering
\includegraphics[width=2.5in]{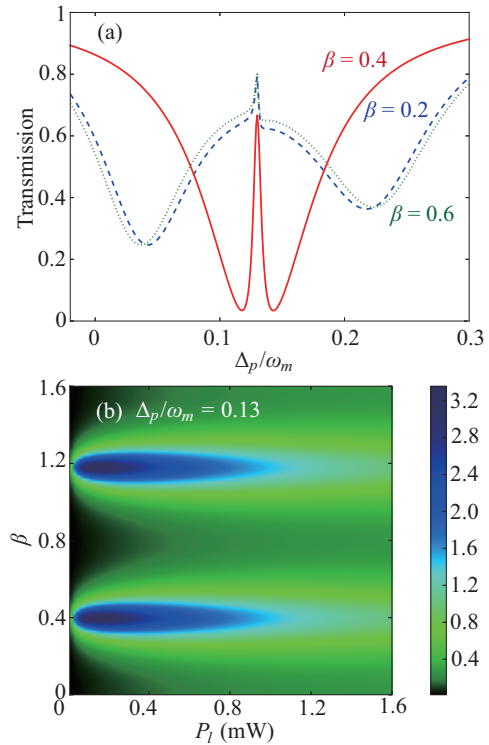}
\caption{(a) Transmission of the probe light as a function of the optical detuning $\Delta_p$. (b) Group delay of the probe light $\tau_g$ (in units of $\mu$s) as a function of the pump power $P_l$ and angle $\beta$. We have selected $P_l=1\,$mW in (a) and $\Delta_a/\omega_m=0.87$ in (a),(b).}
\label{fig4}
\end{figure}

Accompanying the OMIT process, the slow light effect also can emerge \cite{Safavi-Naeini2011,Kippenberg2013}, as characterized by the optical group delay
\begin{align}
\tau_g=\frac{d \arg (t_p)}{d\Delta_p}.
\end{align}
Figure\,\ref{fig3} shows that $\tau_g$ can be tuned by changing the angle $\beta$ and the pump power $P_l$. For $\Delta_p=0$, the group delay can be tuned to be positive or negative, which is clearly different from the case without any particles (i.e., $\tau_g>0$). The means that the nanoparticles can not only able to induce the optical mode shift, but also strongly modify the dispersion of the system. Figure\,\ref{fig3}(b) shows the group delay at $\Delta_p/\omega_m=0.13$, where strong absorption can be achieved by tuning the angle close to the EP. Around the EP ($\beta\approx0.4$ or $1.2$), the fast light can be achieved, corresponding to strong optical absorptions as shown in Fig.\,\ref{fig2}(b).This indicates a new way to achieve slowing or advancing signals by tuning the relative positions of the nanoparticles around a single WGM resonator \cite{Yang2016}.

The nanoparticle-induced frequency shifts weakens the OMIT effect for $\Delta_a/\omega_m=1$; however, this situation can be changed if the effective detuning of the pump is chosen as $$\Delta=\Delta_a+\mathrm{Re}(\epsilon_1+\epsilon_2)=\omega_m.$$ Figure\,\ref{fig4} shows the transmission and group delay of the probe light with $\Delta_a/\omega_m=0.87$. Around the EP (e.g., $\beta\approx 0.4$), an OMIT window emerges at $\Delta_p/\omega_m=0.13$, which is similar to the standard OMIT (i.e., without any nanoparticle), but with an effective detuning of the pump $\Delta/\omega_m\approx1$. Figure\,\ref{fig4}(b) shows that in this case, slow light always exists, which is distinct from the $\Delta_a/\omega_m=1$ case as shown in Fig.\,\ref{fig3}(a).

Our results also reveal that near the EPs, by tuning the pump power, the group delay or advance of the probe can be manipulated in a large scale, i.e., 1.6\,$\mu$s or 2.8\,$\mu$s in Fig.\,\ref{fig3} or Fig.\,\ref{fig4}, respectively. This demonstrates a nontrivial advantage of tuning the system to EPs, where the state and responses of the optomechanical system manifest a high sensitivity in response to the input power, and thereby could be of practical uses in  controlling the propagation of light in COM systems.


\section{Conclusion}

In conclusion, we have studied OMIT in a single resonator coupled with two nanoparticles. We find that by tuning the relative position of the nanoparticles, EPs can emerge periodically and result in significant modifications of both the intracavity optical intensity and the OMIT spectrum. Also, controllable switch from slow to fast light can be achieved in such a system (for other schemes, see e.g., a recent experiment with circuit nanomechanical device \cite{Kippenberg2013}). These results provides a highly sensitive way to tune the light propagation. Besides this, EP-assisted COM devices can also be used for other applications such as EP force sensing \cite{Yang2017,Khajavikhan2017}, topological energy transfer \cite{Harris2016}, and mechanical amplification \cite{Jing2014}.

\section*{ACKNOWLEDGMENTS}
H.L. thanks Ran Huang at Hunan Normal University for useful discussions. H.J. is supported by the HNU Talented Youth Program and the National Natural Science Foundation of China (Grants No. 11474087 and No. 11774086). L.Y. and C.W. are supported by the National Science Foundation under Grant No. EFMA1641109 and the Army Research Office under Grants No. W911NF1210026 and No. W911NF1710189.


\begin{thebibliography}{99}
	
\bibitem{AKM2014}
M. Aspelmeyer, T. J. Kippenberg, and F. Marquardt, Cavity optomechanics, Rev. Mod. Phys. \textbf{86}, 1391 (2014).


\bibitem{Metcalfe2014}
M. Metcalfe, Applications of cavity optomechanics, Appl. Phys. Rev. \textbf{1}, 031105 (2014).


\bibitem{Midolo2018}
L. Midolo, A. Schliesser, and A. Fiore, Nano-opto-electro-mechanical systems, Nat. Nanotechnol. \textbf{13}, 11 (2018).

\bibitem{Bowen2017}
C. Bekker, R. Kalra, C. Baker, and W. P. Bowen, Injection locking of an electro-optomechanical device, Optica \textbf{4}, 1196 (2017).


\bibitem{Polzik2014}
T. Bagci, A. Simonsen, S. Schmid, L. G. Villanueva, E. Zeuthen, J. Appel, J. M. Taylor, A. S\o rensen, K. Usami, A. Schliesser, and E. S. Polzik, Optical detection of radio waves through a nanomechanical transducer, Nature (London) \textbf{507}, 81 (2014).


\bibitem{Teufel2016}
F. Lecocq, J. B. Clark, R. W. Simmonds, J. Aumentado, and J. D. Teufel, Mechanically mediated microwave frequency conversion in the quantum regime, Phys. Rev. Lett. \textbf{116}, 043601 (2016).

\bibitem{Chan2011}
J. Chan, T. P. M. Alegre, A. H. Safavi-Naeini, J. T. Hill, A. Krause, S. Gr\"oblacher, M. Aspelmeyer, and O. Painter, Laser cooling of a nanomechanical oscillator into its quantum ground state, Nature (London) \textbf{478}, 89 (2011).

\bibitem{Teufel2011}
J. D. Teufel, T. Donner, D. Li, J. W. Harlow, M. S. Allman, K. Cicak, A. J. Sirois, J. D. Whittaker, K. W. Lehnert, and R. W. Simmonds, Sideband cooling of micromechanical motion to the quantum ground state, Nature (London) \textbf{475}, 359 (2011).


\bibitem{Wollman2015}
E. E. Wollman, C. U. Lei, A. J. Weinstein, J. Suh, A. Kronwald, F. Marquardt, A. A. Clerk, and K. C. Schwab, Quantum squeezing of motion in a mechanical resonator, Science \textbf{349}, 952 (2015).

\bibitem{Lei2016}
C. U. Lei, A. J. Weinstein, J. Suh, E. E. Wollman, A. Kronwald, F. Marquardt, A. A. Clerk, and K. C. Schwab, Quantum nondemolition measurement of a quantum squeezed state beyond the 3 dB limit, Phys. Rev. Lett. \textbf{117}, 100801 (2016).

\bibitem{Vahala2010}
I. S. Grudinin, H. Lee, O. Painter, and K. J. Vahala, Phonon laser action in a tunable two-level system, Phys. Rev. Lett. \textbf{104}, 083901 (2010).

\bibitem{Jing2014}
H. Jing H, S. K. \"Ozdemir, X. Y. L\"u, J. Zhang, L. Yang, and F. Nori, PT-symmetric phonon laser, Phys. Rev. Lett. \textbf{113}, 053604 (2014).

\bibitem{Arcizet2006}
O. Arcizet, P.-F. Cohadon, T. Briant, M. Pinard, A. Heidmann, J.-M. Mackowski, C. Michel, L. Pinard, O. Français, and L. Rousseau, High-sensitivity optical monitoring of a micromechanical resonator with a quantum-limited optomechanical sensor, Phys. Rev. Lett. \textbf{97}, 133601 (2006).



\bibitem{Kippenberg2012}
E. Gavartin, P. Verlot, and T. J. Kippenberg, A hybrid on-chip optomechanical transducer for ultrasensitive force measurements, Nat. Nanotechnol. \textbf{7}, 509 (2012).





\bibitem{Weis2010}
S. Weis, R. Rivi\`ere, S. Del\'eglise, E. Gavartin, O. Arcizet, A. Schliesser, and T. J. Kippenberg, Optomechanically induced transparency, Science \textbf{330}, 1520 (2010).


\bibitem{Safavi-Naeini2011}
A. H. Safavi-Naeini, T. P. M. Alegre, J. Chan, M. Eichenfield, M. Winger, Q. Lin, J. T. Hill, D. E. Chang, and O. Painter, Electromagnetically induced transparency and slow light with optomechanics, Nature (London) \textbf{472}, 69 (2011).




\bibitem{Kippenberg2013}
X. Zhou, F. Hocke, A. Schliesser, A. Marx, H. Huebl, R. Gross, and T. J. Kippenberg, Slowing, advancing and switching of microwave signals using circuit nanoelectromechanics, Nat. Phys. \textbf{9}, 179 (2013).

\bibitem{Vitali2013}
M. Karuza, C. Biancofiore, M. Bawaj, C. Molinelli, M. Galassi, R. Natali, P. Tombesi, G. Di Giuseppe, and D. Vitali, Optomechanically induced transparency in a membrane-in-the-middle setup at room temperature, Phys. Rev. A \textbf{88}, 013804 (2013).


\bibitem{Shen}
Z. Shen, C.-H. Dong, Y. Chen, Y.-F. Xiao, F.-W. Sun, and G.-C. Guo, Compensation of the Kerr effect for transient optomechanically induced transparency in a silica microsphere, Opt. Lett. \textbf{41}, 1249 (2016).


\bibitem{Xiao2016}
G. Li, X. Jiang, S. Hua, Y. Qin, and M. Xiao, Optomechanically tuned electromagnetically induced transparency-like effect in coupled optical microcavities, Appl. Phys. Lett. \textbf{109}, 261106 (2016).



\bibitem{EIT2005}
M. Fleischhauer, A. Imamoglu, and J. P. Marangos, Electromagnetically induced transparency: optics in coherent media, Rev. Mod. Phys. \textbf{77}, 633 (2005).

\bibitem{Carmon2018}
A. Giorgini, S. Avino, P. Malara, P. De Natale, M. Yannai, T. Carmon, and G. Gagliardi, Stimulated Brillouin cavity optomechanics in liquid droplets, Phys. Rev. Lett. \textbf{120}, 073902 (2018).



\bibitem{Jing2015}
H. Jing, \c{S}. K. \"Ozdemir, Z. Geng, J. Zhang, X.-Y. L\"u, B. Peng, L. Yang, and F. Nori, Optomechanically-induced transparency in parity-time-symmetric microresonators, Sci. Rep. \textbf{5}, 9663 (2015).

\bibitem{Nonlinear}
A. Kronwald and F. Marquardt, Optomechanically induced transparency in the nonlinear quantum regime, Phys. Rev. Lett. \textbf{111}, 133601 (2013).

\bibitem{Liu}
Y. Liu, C. Lu, and L. You, Dissipation-enhanced optomechanically induced transparency, in Conference on Lasers and Electro-Optics, OSA Technical Digest (online) (Optical Society of America, 2018), paper JTh2A.36. 

\bibitem{Cascade}
L. Fan, K. Y. Fong, M. Poot, and H. X. Tang, Cascaded optical transparency in multimode-cavity optomechanical systems, Nat. Commun. \textbf{6}, 5850 (2015).

\bibitem{Xiong2012}
H. Xiong, L. G. Si, A. S. Zheng, X. Yang, and Y. Wu, Higher-order sidebands in optomechanically induced transparency, Phys. Rev. A \textbf{86}, 013815 (2012).


\bibitem{Jing2016}
Y. Jiao, H. L\"u, J. Qian, Y. Li, and Jing H, Nonlinear optomechanics with gain and loss: amplifying higher-order sideband and group delay, New J. Phys. \textbf{18}, 083034 (2016).


\bibitem{Dong2016}
Z. Shen, Y.-L. Zhang, Y. Chen, C.-L. Zou, Y.-F. Xiao, X.-B. Zou, F.-W. Sun, G.-C. Guo, and C.-H. Dong, Experimental realization of optomechanically induced non-reciprocity. Nat. Photonics \textbf{10}, 657 (2016).


\bibitem{Painter2017}
K. Fang, J. Luo, A. Metelmann, M. H. Matheny, F. Marquardt, A. A. Clerk, and O. Painter, Generalized non-reciprocity in an optomechanical circuit via synthetic magnetism and reservoir engineering. Nat. Phys. \textbf{13}, 465 (2017).



\bibitem{spinning}
H. L\"u, Y. Jiang, Y. Z. Wang, H. Jing, Optomechanically induced transparency in a spinning resonator, Photonics Res. \textbf{5}, 367 (2017).

\bibitem{Painter2012}
J. T. Hill, A. H. Safavi-Naeini, J. Chan, and O. Painter, Coherent optical wavelength conversion via cavity optomechanics, Nat Commun. \textbf{3}, 1196 (2012).




\bibitem{Wu2017}
H. Xiong, Z. X. Liu, and Y. Wu, Highly sensitive optical sensor for precision measurement of electrical charges based on optomechanically induced difference-sideband generation, Opt. Lett. \textbf{42}, 3630 (2017).


\bibitem{Bender1998}
C. M. Bender and S. Boettcher, Real spectra in non-Hermitian Hamiltonians having PT symmetry, Phys. Rev. Lett. \textbf{80}, 5243 (1998).


\bibitem{Bender2007}
C. M. Bender, Making sense of non-Hermitian Hamiltonians, Rep. Prog. Phys. \textbf{70}, 947 (2007).

\bibitem{PT2016}
V. V. Konotop, J. Yang, and D. A. Zezyulin, Nonlinear waves in PT-symmetric systems. Rev. Mod. Phys. \textbf{88}, 035002 (2016).

\bibitem{Feng2017PT}
L. Feng, R. El-Ganainy, and L. Ge, Non-Hermitian photonics based on parity-time symmetry, Nat. Photonics \textbf{11}, 752 (2017).

\bibitem{Peng2014}
B. Peng, \c{S}. K. \"Ozdemir, F. Lei, F. Monifi, M. Gianfreda, G. L. Long, S. Fan, F. Nori, C. M. Bender, and L. Yang, Parity-time-symmetric whispering-gallery microcavities, Nat. Phys. \textbf{10}, 394 (2014).

\bibitem{Ruter2014}
C. E. R\"uter, K. G. Makris, R. El-Ganainy, D. N. Christodoulides, M. Segev, and D. Kip, Observation of parity-time symmetry in optics, Nat. Phys. \textbf{6}, 192 (2010).

\bibitem{Zyablovsky}
A. A. Zyablovsky, A. P. Vinogradov, A. A. Pukhov, A. V. Dorofeenko, and A. A. Lisyansky, PT-symmetry in optics, Phys. Usp. \textbf{57}, 1063 (2014).


\bibitem{Longhi2018}
S. Longhi, Parity-time symmetry meets photonics: a new twist in non-Hermitian optics, Europhys. Lett. \textbf{120}, 64001 (2018).


\bibitem{Guo2009}
A. Guo A, G. J. Salamo, D. Duchesne D,  R. Morandotti, M. Volatier-Ravat, V. Aimez, G. A. Siviloglou, and D. N. Christodoulides, Observation of PT-symmetry breaking in complex optical potentials, Phys. Rev. Lett. \textbf{103}, 093902 (2009).


\bibitem{Peng2014Science}	
B. Peng, \c{S}. K. \"{O}zdemir, S. Rotter, H. Yilmaz, M. Liertzer, F. Monifi, C. M. Bender, F. Nori, and L. Yang, Loss-induced suppression and revival of lasing, Science \textbf{346}, 328 (2014).

\bibitem{Feng2017}
S. Longhi and L. Feng, Unidirectional lasing in semiconductor microring lasers at an exceptional point, Photonics Res. \textbf{5}, B1 (2017).


\bibitem{Invisibility}
Z. Lin, H. Ramezani, T. Eichelkraut, T. Kottos, H. Cao, and D. N. Christodoulides, Unidirectional invisibility induced by PT-symmetric periodic structures, Phys. Rev. Lett. \textbf{106}, 213901 (2011).

\bibitem{Fan2017}
S. Assawaworrarit, X. Yu, and S. Fan, Robust wireless power transfer using a nonlinear parity-time-symmetric circuit, Nature (London) \textbf{546}, 387 (2017).


\bibitem{Zhen2018}
H. Zhou, C. Peng, Y. Yoon, C. W. Hsu, K. A. Nelson, L. Fu, J. D. Joannopoulos, M. Solja\v{c}i\'c, and B. Zhen, Observation of bulk Fermi arc and polarization half charge from paired exceptional points, Science \textbf{359}, 1009 (2018).


\bibitem{YangBiao2018}
B. Yang, Q. Guo, B. Tremain, R. Liu, L. E. Barr, Q. Yan, W. Gao, H. Liu, Y. Xiang, J. Chen, C. Fang, A. Hibbins, L. Lu, and S. Zhang, Ideal Weyl points and helicoid surface states in artificial photonic crystal structures, Science \textbf{359}, 1013 (2018).


\bibitem{Harris2016}
H. Xu, D. Mason, L. Jiang, and J. G. E. Harris, Topological energy transfer in an optomechanical system with exceptional points, Nature (London) \textbf{537}, 80 (2016).

\bibitem{Jing2017}
H. L\"u, S. K. \"Ozdemir, L. M. Kuang, F. Nori, and H. Jing, Exceptional points in random-defect phonon lasers, Phys. Rev. Applied \textbf{8}, 044020 (2017).

\bibitem{Jing2017cooling}
H. Jing, \c{S}. K. \"Ozdemir, H. L\"u, and F. Nori, High-order exceptional points in optomechanics, Sci. Rep. \textbf{7}, 3386 (2017).

\bibitem{Renault2017}
P. Renault, H. Yamaguchi, and I. Mahboob, PT-symmetry breaking in non-identical electromechanical resonators, arXiv: 1708.02352.

\bibitem{Yang2016}
B. Peng, \c{S}. K. \"Ozdemir, M. Liertzer, W. Chen, J. Kramer, H. Y\i lmaz, J. Wiersig, S. Rotter, and L. Yang, Chiral modes and directional lasing at exceptional points, Proc. Natl. Acad. Sci. USA \textbf{113}, 6845 (2016).


\bibitem{Yang2017}
W. Chen, \c{S}. K. \"Ozdemir, G. Zhao, J. Wiersig, and L. Yang, Exceptional points enhance sensing in an optical microcavity, Nature (London) \textbf{548}, 192 (2017).

\bibitem{Wiersig2014}	
J. Wiersig, Enhancing the sensitivity of frequency and energy splitting detection by using exceptional points: application to microcavity sensors for single-particle detection, Phys. Rev. Lett. \textbf{112}, 203901 (2014).	

\bibitem{Wiersig2016}
J. Wiersig, Sensors operating at exceptional points: general theory, Phys. Rev. A \textbf{93}, 033809 (2016).






\bibitem{Clerk2016}
M. A. Lemonde, N. Didier, and A. A. Clerk, Enhanced nonlinear interactions in quantum optomechanics via mechanical amplification, Nat. Commun. \textbf{7}, 11338 (2016).

\bibitem{Verhagen2017}
R. Leijssen, G. R. La Gala, L. Freisem, J. T. Muhonen, and E. Verhagen, Nonlinear cavity optomechanics with nanomechanical thermal fluctuations, Nat. Commun. \textbf{8}, 16024 (2017).

\bibitem{Marquardt2015}
V. Peano, C. Brendel, M. Schmidt, and F. Marquardt, Topological phases of sound and light, Phys. Rev. X \textbf{5}, 031011 (2015).



\bibitem{Yang2010}
J. Zhu, \c{S}. K. \"Ozdemir, L. He, and L. Yang, Controlled manipulation of mode splitting in an optical microcavity by two Rayleigh scatterers, Opt. Express \textbf{18}, 23535 (2010).




\bibitem{Khajavikhan2017}
H. Hodaei, A. U. Hassan, S. Wittek, H. Garcia-Gracia, R. El-Ganainy, D. N. Christodoulides, and M. Khajavikhan, Enhanced sensitivity at higher-order exceptional points, Nature (London) \textbf{548}, 187 (2017).











\end{thebibliography}
\end{document}